\documentclass[manuscript]{acmart}
\AtBeginDocument{%
  \providecommand\BibTeX{{%
    \normalfont B\kern-0.5em{\scshape i\kern-0.25em b}\kern-0.8em\TeX}}}

\copyrightyear{2021} 
\acmYear{2021} 
\setcopyright{acmcopyright}\acmConference[ARES 2021]{The 16th International Conference on Availability, Reliability and Security}{August 17--20, 2021}{Vienna, Austria}
\acmBooktitle{The 16th International Conference on Availability, Reliability and Security (ARES 2021), August 17--20, 2021, Vienna, Austria}
\acmPrice{15.00}
\acmDOI{10.1145/3465481.3470052}
\acmISBN{978-1-4503-9051-4/21/08}

\usepackage{array}

\begin{document}

\title{Risk-Oriented Design Approach For Forensic-Ready Software Systems}

\author{Lukas Daubner}
\email{daubner@mail.muni.cz}
\orcid{0000-0003-0853-2776}
\affiliation{
  \institution{Faculty of Informatics, Masaryk University}
  \city{Brno}
  \country{Czech Republic}}

\author{Raimundas Matulevi\v{c}ius}
\email{raimundas.matulevicius@ut.ee}
\orcid{0000-0002-1829-4794}
\affiliation{
  \institution{University of Tartu}
  \city{Tartu}
  \country{Estonia}}

\renewcommand{\shortauthors}{Daubner and Matulevi\v{c}ius}

\begin{abstract}
Digital forensic investigation is a complex and time-consuming activity in response to a cybersecurity incident or cybercrime to answer questions related to it. These typically are what happened, when, where, how, and who is responsible. However, answering them is often very laborious and sometimes outright impossible due to a lack of useable data. The forensic-ready software systems are designed to produce valuable on-point data for use in the investigation with potentially high evidence value. Still, the particular ways to develop these systems are currently not explored.

This paper proposes consideration of forensic readiness within security risk management to refine specific requirements on forensic-ready software systems. The idea is to re-evaluate the taken security risk decisions with the aim to provide trustable data when the security measures fail. Additionally, it also considers possible disputes, which the digital evidence can solve. Our proposed approach, risk-oriented forensic-ready design, composes of two parts: (1) process guiding the identification of the requirements in the form of potential evidence sources, and (2) supporting BPMN notation capturing the potential evidence sources and their relationship. Together they are aimed to provide a high-level overview of the forensic-ready requirements within the system. Finally, the approach is demonstrated on an automated valet parking scenario, followed by a discussion regarding its impact and usefulness within the forensic readiness effort.
\end{abstract}




\maketitle
\section{Introduction}
With the increasing importance and ubiquity of digital technology in our daily life, so is the danger of cyberattacks. The vast spectrum of attack targets includes smart home devices, vehicles, industrial systems, and electrical grids~\cite{Stellios:2018}. For this reason, protection against cyberattacks and designing of secure systems is nowadays an important topic~\cite{Geismann:2020}. However, despite the employed measures, the possibility of a cyberattack cannot be ruled out, and with it arises the need to gain insight into its occurrence~\cite{Casey:2020}.

Digital forensics should answer what happened, when, where, how, and who is responsible. It provides the methods to investigate by collecting, analyzing, reconstructing a chain of events, and finally presenting the answers based on artefacts called digital evidence~\cite{Casey:2011}. In contrast to the reactive nature of the investigation, proactive measures, called forensic readiness, are employed to maximize the usefulness of incident evidence data and minimize the cost of forensics during an incident response~\cite{Tan:2001}. It can also be seen as a natural enhancement of security practices.

An approach to forensic readiness from the software engineering perspective introduced forensic-ready software systems~\cite{Pasquale:2018}. Essentially, such systems can produce sufficient, relevant, trustworthy, and non-repudiable digital evidence that is handled in a forensically sound way and conforming to legal obligations. This approach is in contrast to the organization-centric ones commonly taken to achieve forensic readiness~\cite{Rowlingson:2004}. However, one should consider both approaches simultaneously while also taking into account and applying existing security practices.

There are many gaps in the designing of forensic-ready software systems. While the broad set of forensic-ready requirements has been formulated, the relevant requirements engineering~\cite{Sommerville:2010} approaches are not yet established. For example, the availability of evidence is an important requirement. However, the means to discover, organize, prioritize, and specify the evidence sources of digital evidence within the system are not present.

The goal of this paper is to introduce an approach to designing forensic-ready software systems. This approach focuses on the capability to produce potential digital evidence\footnote{Note the difference: potential digital evidence --- potentially useable for future investigation, and digital evidence --- used to satisfy or refute the investigation hypothesis.}. Specifically, it follows the risk-oriented approach, building on and expanding a well-matured security risk management~\cite{Matulevivcius:2017}. This novel Risk-Oriented Forensic-Ready Design approach is demonstrated on a running scenario on which the motivation, process, supporting BPMN notation, and interplay with security risk is shown.

This paper is organized as follows. Section~\ref{sec:background} outlines the related work. Section~\ref{sec:runningScenario} presents the running scenario and considered security attacks, and discusses initial forensic-ready capabilities. Section~\ref{sec:approach} describes the approach with both its components, process and notation. It is followed by Section~\ref{sec:demo}, which demonstrates the application of the previously described approach to the scenario. Section~\ref{sec:discussion} then discusses the capabilities, improvements, and possible evolutions of the approach. Finally, Section~\ref{sec:conclusion} concludes the paper.

\section{Background}
\label{sec:background}



Several frameworks focusing on forensic-ready software systems, sometimes called forensic-by-design, were formulated, focusing on a specific domain~\cite{AbRahman:2016, Grispos:2017b}. In the time-frame of acquiring the general idea of approaching forensic readiness from a software engineering perspective~\cite{Pasquale:2018}, more works started to appear. An example is the preservation of minimal and relevant digital evidence~\cite{Alrajeh:2017} and automated logging instrumentation~\cite{RiveraOrtiz:2020}, together with proposals on verification methods of such systems~\cite{Daubner:2020a}.

Concrete design methods for forensic-ready software systems are largely unexplored. However, the related topic of designing secure systems is well-established. Both forensic-ready and security concerns have significant overlap~\cite{Grobler:2007} and can be considered as high-level non-functional requirements. As such, they need to be addressed at the architectural level~\cite{Chung:2012}. Within the designing of secure systems, a prime approach is UMLsec~\cite{Jurjens:2002}, which captures security concerns into UML models. Another example is SEED~\cite{Vasilevskaya:2015}, which emphasizes the separation of concerns.


\textbf{Security risk management} is a well-known approach to the identification of security requirements by assessing risks posed to the systems~\cite{Matulevivcius:2017}. Following the model-based approach, the domain model of information systems security risk management (ISSRM) has been formulated~\cite{Mayer:2009}, together with a process to manage the security risks. With the ISSRM domain model as a basis, a modelling approach based on the business process model and notation (BPMN) was formulated~\cite{Altuhhova:2013}, which was later utilized to define risk-oriented security patterns~\cite{Ahmed:2014}. The other modelling approaches aimed at the security risk are Secure Tropos~\cite{Matulevicius:2008}, Mal-Activities~\cite{Chowdhury:2012}, and Misuse Cases~\cite{Soomro:2013}.

The risk-based approach was identified as a recommended strategy to approach forensic readiness~\cite{Kazadi:2015}. Elsewhere in~\cite{AbRahman:2016, Grispos:2017b}, the forensic readiness frameworks also highlight that risk management is an important factor. Furthermore, the security risk is also described as a driving force in establishing forensic readiness in smart buildings~\cite{Bajramovic:2016}.

\section{Running Scenario}
\label{sec:runningScenario}

This section introduces the running scenario, an Automated Valet Parking (AVP), to demonstrate our proposed approach. The scenario represented by a BPMN model has been already utilized for the application of privacy-enhancing technologies~\cite{Nwaokolo:2020}. We will describe the basic scenario with its innate forensic capabilities and weaknesses demonstrated by four attack scenarios. The particular weaknesses are then used as a motivation for defining a process to improve forensic readiness, supporting notation, and enhanced forensic-ready scenario in later sections.

\begin{figure*}
    \centering
    \begin{minipage}{\linewidth}
        \begin{center}
        \includegraphics[scale=.33]{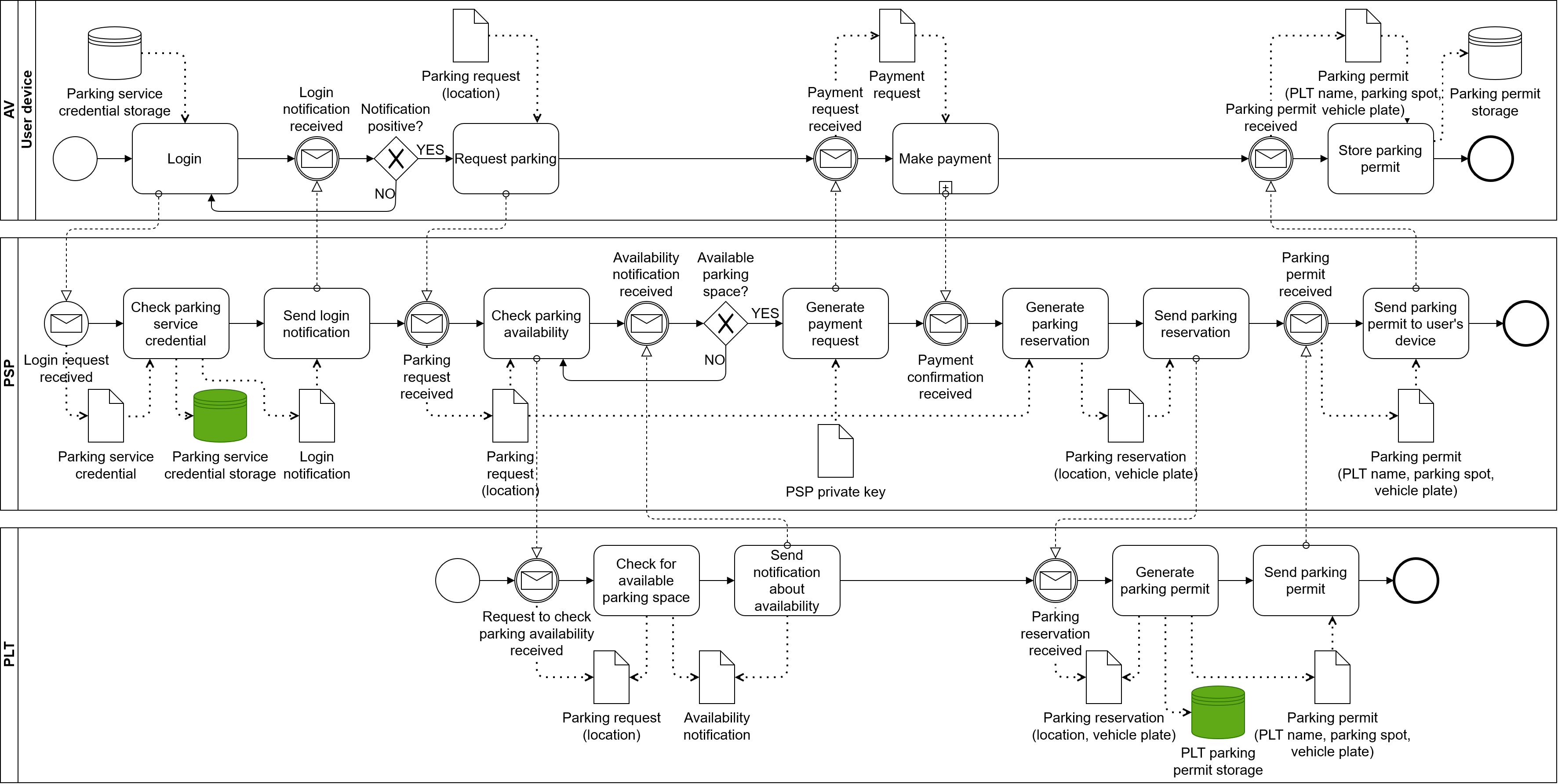}
        \end{center}
        \footnotesize
        \quad \emph{Note:} Innate sources of potential forensic evidence sources are highlighted in green.
    \end{minipage}
    \caption{Automated Valet Parking Scenario: Issuing a Permit}
    \label{fig:scenarioBasic}
\end{figure*}

\subsection{Default Process Model}

The AVP is a service enabling user to leave an autonomous vehicle in a drop-off area to park it automatically, without further assistance. The vehicle would connect to a parking lot and self-drive to a designated vacant parking space~\cite{Ni:2019}. An example of an AVP implementation is a joint project by Mercedes-Benz and Bosh~\cite{Daimler:2020}.

In this paper, a part of the AVP process is considered and modelled using BPMN. Concretely, the focused part is on the process of issuing a parking permit. Figure~\ref{fig:scenarioBasic} contains a model of this process in BPMN. The model is focusing on the business aspect, not considering the security nor forensic readiness. However, such a model can be used as a starting point for a security and forensic-ready risk analysis as it considers the impacts on business.

The following observations can be made, which affects reasoning about the potential digital evidence:
\begin{itemize}
    \item \textbf{Autonomous Vehicle (AV)} –-- A user device initiating the process and out of stakeholder control. The would-be potential evidence within this entity is not useable, as user cooperation cannot be expected during any investigation.
    \item \textbf{Parking Service Provider (PSP)} –-- A service that mediates exchange between AVs and PLTs. Responsible for handling parking reservations, payment requests, payment confirmations, and parking permits. Fully under stakeholder control. 
    \item \textbf{Parking Lot Terminal (PLT)} –-- An IoT device is controlling access to the parking lot by issuing parking permits. It serves as an authoritative source of parking lot state. Under stakeholder control. Potentially easy physical access.
\end{itemize}
Based on this enumeration, the sources of potential evidence are highlighted in Figure~\ref{fig:scenarioBasic}. Both entities are persistent and can be modified by the process run. Thus, being an effect (creating a parking permit) of a cause (receiving parking reservation). However, the level of certainty and sufficiency of such a causal relationship needs to be established.

\subsection{Attack Scenarios}

Based on the default process model, four attack scenarios demonstrate the weaknesses and insufficiency of the innate evidence sources. Specifically, each attack scenario's occurrence must yield an indicator of compromise, which would differentiate it from a nominal run. To reason about the scenarios, they are modelled by using Security Risk-aware BPMN~\cite{Altuhhova:2013} models. Each of the models represents a risk model corresponding to the particular attack scenario.

\paragraph{Attack Scenario \#1: Falsifying Availability Reports}

In this scenario, an attacker can intercept a request to a PLT and send a negative reply, regardless of the actual parking space availability and ability of PLT to process the request. The disruption is performed at random times, and when it is not, the process resumes in normal behaviour. Given the random nature of the attack, it is hard to detect while still impacting the service availability leading to a financial loss.

\begin{figure}
    \centering
    \includegraphics[scale=0.35]{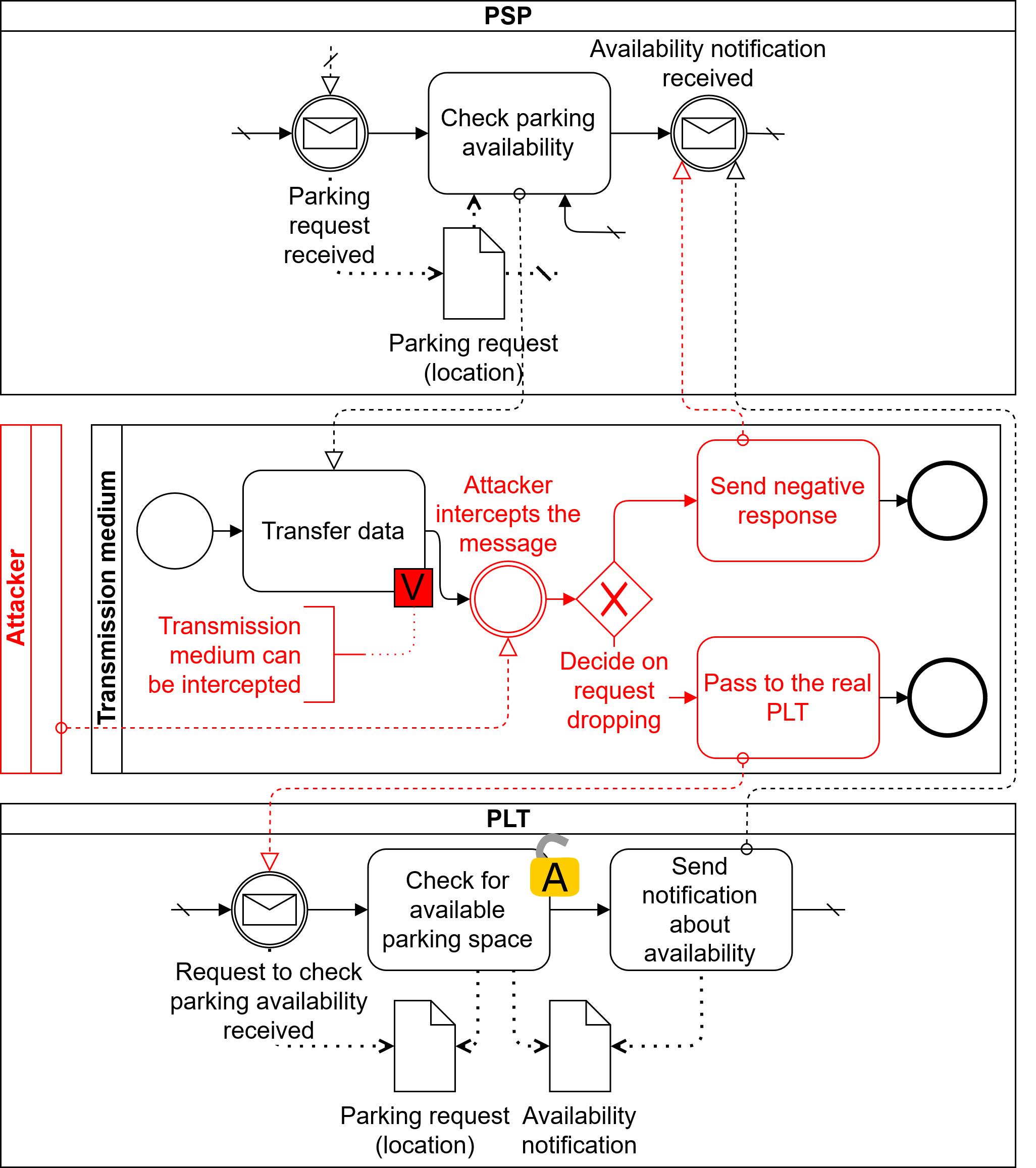}
    \caption{Attack Scenario \#1: Falsifying Availability Reports}
    \label{fig:attack1}
\end{figure}

The affected part of a model is illustrated in Figure~\ref{fig:attack1}. In this case, none of the innate potential evidence sources is involved, and all data involved are not persistent. Thus, there is no record (an indicator of compromise) of the attack or abnornal behaviour.

\paragraph{Attack Scenario \#2: Fabricating a Parking Permit}

\begin{figure}
    \centering
    \includegraphics[scale=0.35]{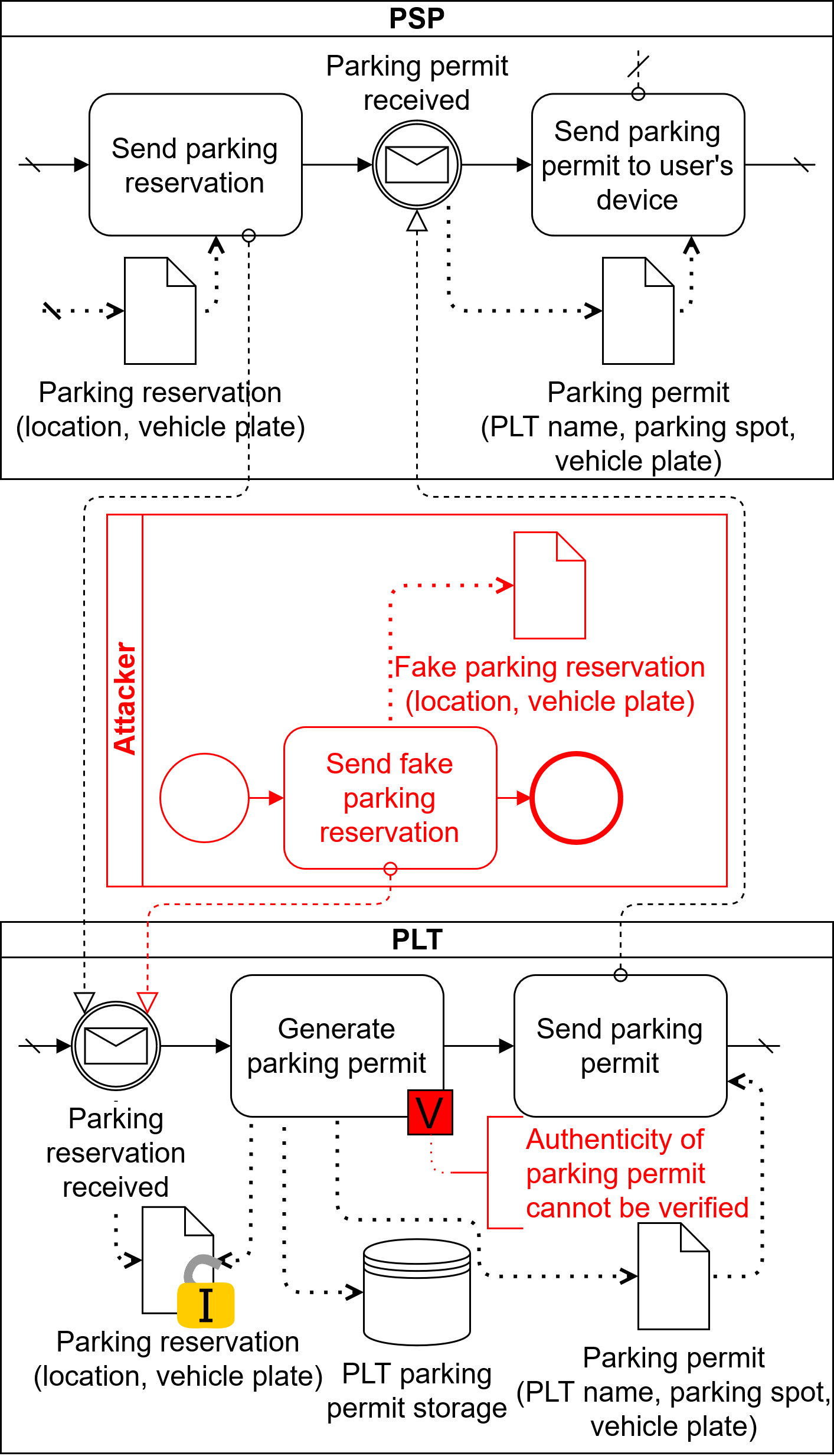}
    \caption{Attack Scenario \#2: Fabricating a Parking Permit}
    \label{fig:attack2}
\end{figure}

In this scenario, an attacker bypasses the PSP and sends a forged parking reservation directly to PLT, which generates a valid parking permit. A successful attack completely avoids payment, thus inducing financial loss. Figure~\ref{fig:attack2} contains a model illustrating one of the possible ways of commencing the attack. Another way can be a direct injection of a parking permit into the parking permit store off-process.

In contrast with the previous scenario, the potential evidence source is affected as the falsified parking permit must be inserted into the storage. It should be noted that the parking permit must contain an actual vehicle plate number to be valid, possibly identifying the assailant. However, it is impossible to determine whether the parking permit is genuine or falsified only with the storage. Consequently, parking permit storage alone is not a sufficient indicator of compromise.

\paragraph{Attack Scenario \#3: Issuing Fake Payment}

\begin{figure}
    \centering
    \includegraphics[scale=0.35]{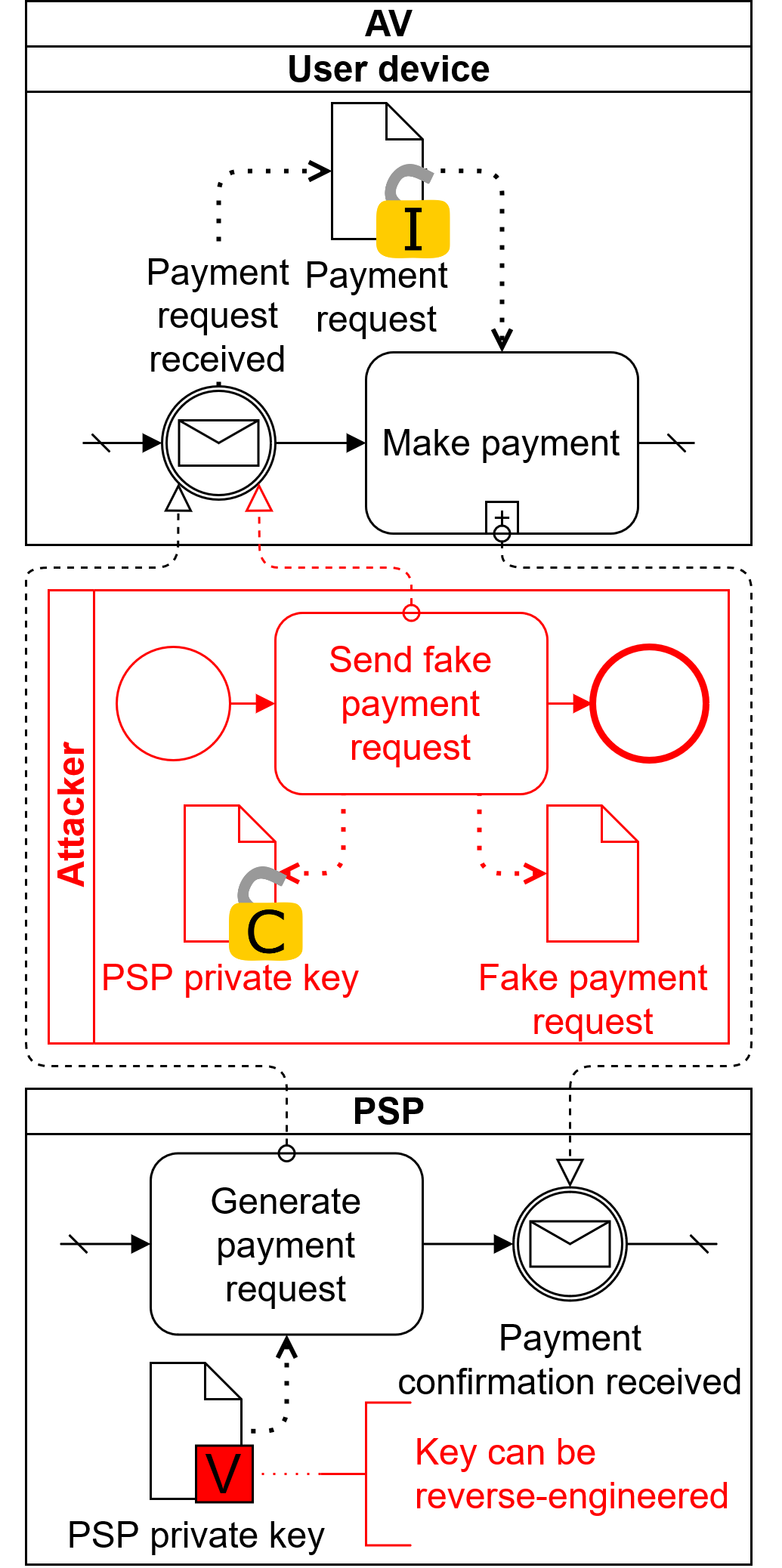}
    \caption{Attack Scenario \#3: Issuing Fake Payment}
    \label{fig:attack3}
\end{figure}

In this scenario, an attacker impersonates PSP and issues a payment request. Such an attack can lead to both financial losses due to potential reimbursement of affected customers, as well as loss of reputation. The corresponding risk model is illustrated in Figure~\ref{fig:attack3}.

Locating potential evidence source is especially challenging in this case because the user device is the primary target. However, even with a willing user providing their data, there is no persistent data that could be used as evidence. PSP also does not store the payment requests or the payment confirmations, so an indicator that the request was indeed malicious is impossible. Certainly, out-of-process evidence like the account statement could potentially yield critical information, but the connection to PSP still cannot be decided.

\paragraph{Attack Scenario \#4: Repudiation of a Parking Permit}

\begin{figure}
    \centering
    \includegraphics[scale=0.35]{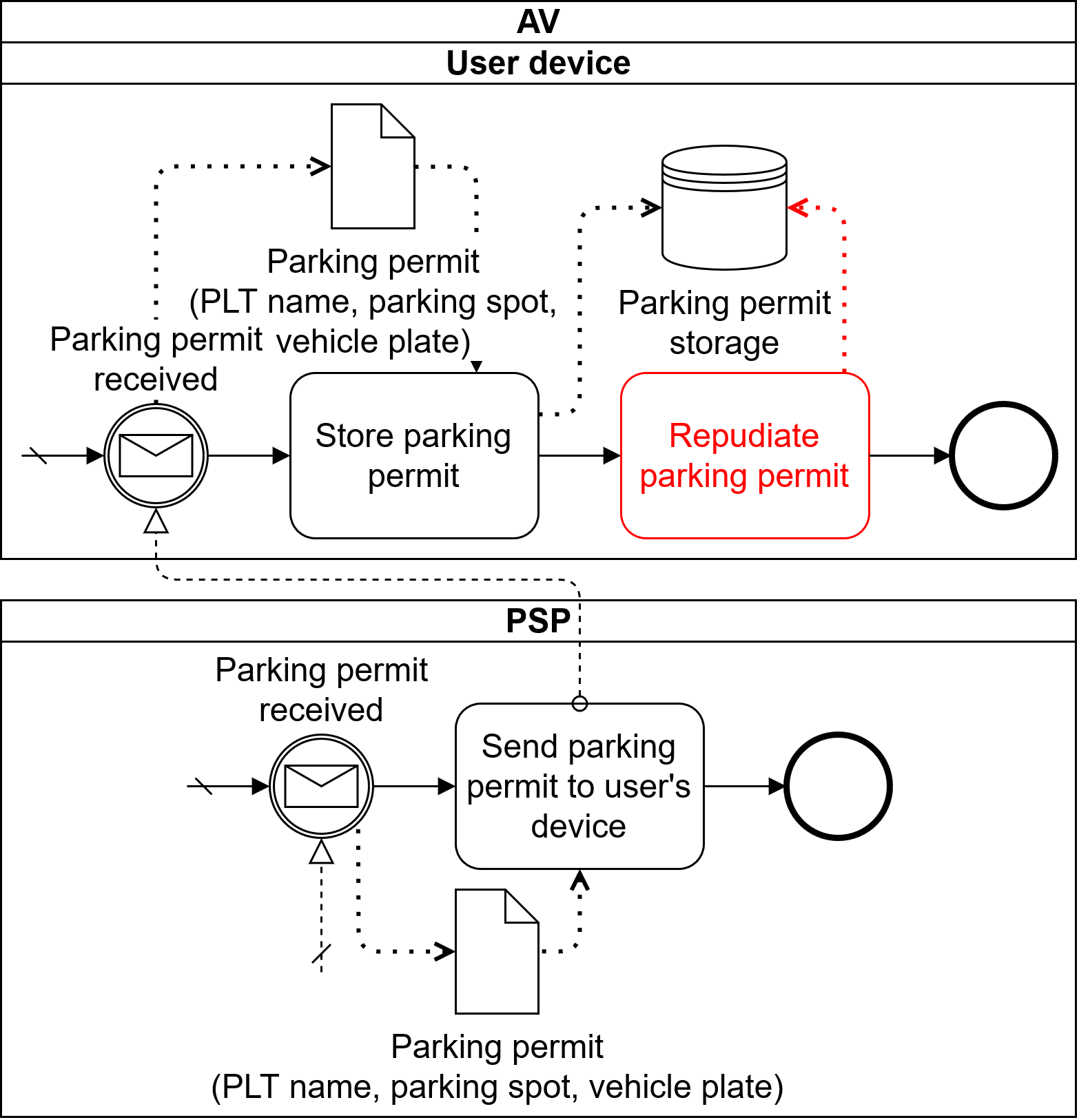}
    \caption{Attack Scenario \#4: Repudiation of a Parking Permit}
    \label{fig:attack4}
\end{figure}

This scenario is an example of an atypical attack, for which forensic readiness is a good countermeasure. Specifically, after the nominal execution of the process, a dishonest user deletes and repudiates an unused parking permit to get reimbursement with a claim that they did not receive it. Figure~\ref{fig:attack4} illustrates the part of the model in question.

The main obstacle is that the malicious behaviour is happening on the user device, outside the stakeholders reach. In this case, the parking permit storage shows the generation of the corresponding parking permit, thus resulting in a piece of evidence. However, the dishonest user might claim that the parking permit was not delivered due to a bug or planted after the complaint. In the default setting, it cannot be proven otherwise.

\subsection{Discussion}

Summarizing the insights from all four attack scenarios, it is clear that the innate sources of potential evidence produce insufficient data for possible investigation. The only usable potential evidence source in the default process model is the Parking permit storage. However, it can be disputed as the transmission of the parking permit to AV is not proved, and there is no other data to support it. Consequently, neither nominal nor anomalous process execution cannot be traced.

Additionally, low-level evidence sources are, in fact, present even in the default setting. The execution of the process within the information system will very probably generate background data. For example, communication between PSP and AV will likely produce a log record by a web server, firewall, or proxy. Such data are in practice widely utilized during the forensic investigation~\cite{Chuvakin:2012}. However, their discovery is typically performed in an ad-hoc way~\cite{Casey:2011}, they are not made for the forensic task in the first place, and their complexity, variety, and volume poses a significant challenge~\cite{Garfinkel:2010}. Therefore, proper introduction, and consequently documentation, of the potential evidence sources within the process eases future investigation and makes an effective indicator of compromise~\cite{Daubner:2020b}.

Forensic readiness is tightly related to security. In many cases, implementing security controls could mitigate the presented attack scenarios. For example, employing public-key cryptography mitigates attack scenario \#1. However, there is a risk of reverse-engineering the PLT private key, which would re-enable the attack. The core difference between forensic readiness and security is that while security aims to prevent the incident, forensic readiness is about having detailed data about the incident occurring and its context. Essentially, forensic readiness provides valuable data when security measures fail.

Due to the close relation, implementing the forensic readiness capabilities is best done in a supporting role to the security. Particularly interesting for forensic readiness is building upon security risk management, a systematical approach to security objectives, risks, and risk treatment. Specifically, any leftover security risk that was retained is an excellent target to cover by inserting the forensic-ready measures. A good example is threats by insiders that could misuse the secure information system both maliciously and unintentional as they have valid access to it. Arguably, the forensic readiness capabilities can also provide data for risks unconsidered during the design. For those reasons, we propose risk-oriented forensic-ready analysis as a complementary effort to security risk management.

\section{Risk-Oriented Forensic-Ready Design}
\label{sec:approach}

Our proposed method consists of two major components: Process and Modelling. The first part is based upon the Information Systems Security Risk Management (ISSRM) process~\cite{Mayer:2009}, extending it with the inclusion of forensic readiness concerns. The second part supports the process by introducing forensic-ready notation in the BPMN modelling language.

\subsection{Forensic-Ready Risk Analysis Process}

The core of the forensic-ready risk analysis method is an iterative process aimed at the identification of potential evidence sources and covering residual risk from security risk analysis. Additionally, it directs the analysts to discover new unaccounted risks and consider precautions relevant in establishing a forensic-ready information system.
\begin{figure}
    \centering
    \includegraphics[scale=0.9]{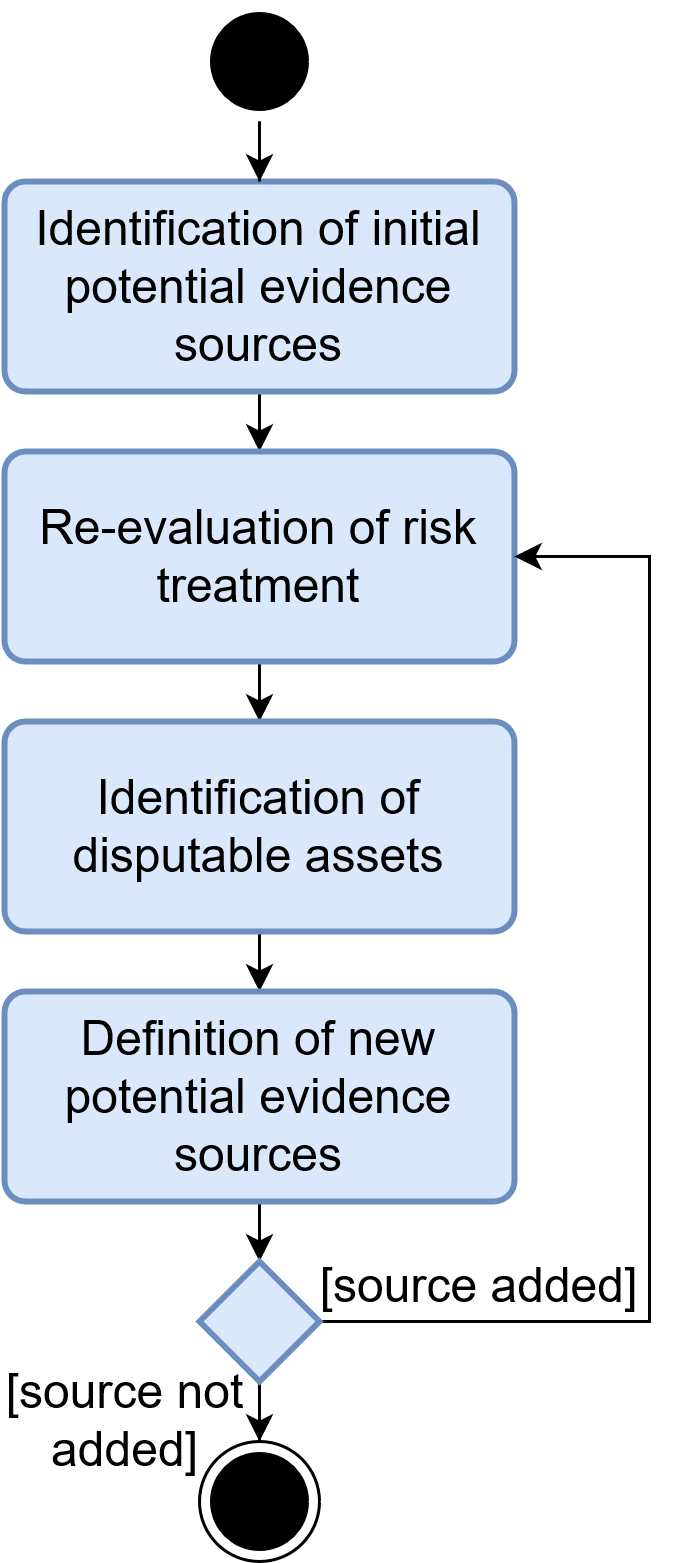}
    \caption{Forensic-Ready Risk Analysis Process}
    \label{fig:process}
\end{figure}

The presented forensic-ready risk analysis process is an extension to already established security risk analysis. It assumes that the security risk was completed and builds on its results. Figure~\ref{fig:process} illustrates the forensic-ready risk analysis process, with the individual steps that are further described in the following paragraphs. Generally, the steps follow the well-known guidelines formulated by Rowlingson~\cite{Rowlingson:2004} relevant to the context of information systems.

\paragraph{Identification of initial potential evidence sources} Firstly, we introduce a new entity within the ISSRM domain model, the \textit{Potential evidence source}. It is a combination of a business asset representing the data, and a supporting IS asset representing the relevant part of the information system generating it. The responsibility of the possible evidence source is to produce instances of potential evidence that within an evidence context records the occurrence of some particular event or sequence of events. Eliciting the existing potential evidence is vital in judging the needs for the implementation of further controls to have sufficient records of the events occurring in the system to support or refute an investigation hypothesis.

Initially, the main focus should be on the sources with business asset data (i.e. background evidence), followed by the sources based on the implemented security controls (i.e. foreground evidence). Sources of the latter are the IS assets like logs, monitoring, or IDS. Additionally, broader scope of the sources like records from CCTV cameras or RFID access logs should be included, if applicable. 

Each identified potential evidence source should be described as detailed as possible. Primarily, each type of data produced, including its content, format, and properties of its storage. The important storage-related information is location, access control, and retention. Such description is useful both for the subsequent steps but also for the potential investigation.

\paragraph{Re-evaluation of risk treatment} For this step, it is crucial to building on the assumption that security risk analysis was already performed and the overall taken risk treatment are deemed satisfactory. The decisions are reviewed whether there is a need for an additional treatment, ultimately resulting in new potential evidence sources. A further consideration is placed on the risks currently unknown, i.e., they are present but not identified by the risk analysis at the time. The risk can be categorized to Retained risk, Residual risk, and Unknown risk.

\textbf{Retained risk} is the type of risk coming from security risk analysis with risk retention decision, i.e., it has been accepted with no further action. We recommend re-evaluating the decisions whether it is feasible to collect data showing the occurrence of its associated event and the actual impact.

\textbf{Residual risk} builds on the assumption that it is usually practically impossible to mitigate security risk fully. It can be understood as a risk when deciding to reduce or transfer it, still having a non-zero probability of event or negative consequence. Forensic-ready treatment might be feasible to detect its occurrence and gather additional data about it.

\textbf{Unknown risk} is a risk that is not identified at the time. It includes undetected, zero-day vulnerabilities and simply unconsidered or overlooked items in the risk analysis. Indeed, it very hard and even impossible to mitigate such risks before being aware of their existence. However, forensic-ready treatment can help improve the odds of detecting and subsequently investigating an occurrence of a previously unknown threat. In general, detecting the unknown includes identifying indicators of compromise, having a notion about the good and bad system behaviour, and making possible attacker noisy by producing potential evidence at the key places.

The iterative nature of the process giving a way to evaluate the effect of forensic-ready treatment, which means the newly induced controls in the form of potential evidence sources, whether they sufficiently cover the risk at hand. It is important to note that the forensic-ready treatment will not reduce nor transfer the risk. However, it enables one to be aware of its occurrence and possibly have enough evidence to assess the impact or prosecute the assailant.

\paragraph{Identification of disputable assets} Forensic readiness does not only complement the security, but it also brings about precautions to business threats. As identified in Rowlingston’s guidelines~\cite{Rowlingson:2004}, the potential evidence can also be used to support internal disciplinary issues and disputes. Furthermore, they can demonstrate damage of assets, outages, faults to prove a financial loss or breach of contract.

In the third step, the analyst should assess the risk of dispute over some asset. The examples of dispute types are (non-)existence, accuracy, tampering, mishandling, outage, and fault. Again, the taken forensic-ready controls should prove the occurrence of some event and its impact.

Other potential evidence data might also be a target of dispute. It is especially true during the investigation, where the integrity of evidence must be recorded. However, such measures applied proactively might be required if there is a high risk of dispute. The goal is to improve the overall value of the potential evidence. Generally, there are three categories of additional measures for potential evidence sources:
\begin{itemize}
    \item \textbf{Precision} –-- Increase the accuracy of its data content.
    \item \textbf{Protection} –-- Create a proof of integrity, authenticity, or non-repudiation.
    \item \textbf{Enhancement} –-- Add additional, detailed information to the data or documentation
\end{itemize}

\paragraph{Definition of new potential evidence sources} The last step is to elicit new potential evidence sources based on the decisions from the previous step. Each new potential evidence source must define the IS asset which generates it together with the data. The definition of a new source corresponds to the formulation and implementation of new requirements in the system. Furthermore, a possible relationship between the data from distinct potential evidence sources should be recorded. That means both dependence of one data to another and strengthening coming from the treatment of disputable data.

Analogously to the first step that identified initial potential evidence sources, the new ones should also contain information about content, format, and storage. The process terminates when one does not add any new evidence source. Otherwise, it continues with another iteration, which might introduce another risk or dispute.

\subsection{Forensic-Ready BPMN Notation}

We demonstrate the modelling notation for the risk-oriented forensic-ready design to support the described process and enable reasoning over the potential evidence and risks. We extend the BPMN modelling language 2.0~\cite{Silver:2009} at its descriptive modelling level. Furthermore, the notation is designed to complement the Security Risk-Aware BPMN~\cite{Altuhhova:2013} in the same way as the process complement the security risk analysis. Even though the BPMN is not primarily focused on security nor forensic concerns, it allows a simple, high-level view.

The notation is designed to assist in the following ways:
\begin{itemize}
    \item Captures initially and subsequently defined potential evidence sources. Both cases are modelled with the explicit association on their data types, meaning a Data Object representing the Business asset. The same approach applies to the security risk and for possibly disputed actions.
    \item Assists in the re-evaluation of risk treatment by putting together potential evidence sources and Security Risk-Aware BPMN risk-related constructs~\cite{Altuhhova:2013}. The alignment of the potential evidence sources, their data types, and their relationships forms a context to capture specific events.
    \item Supports the introduction of protective measures on potential evidence source data as a relationship with another potential evidence source data that is strengthening it and essentially provides proof of the former.
\end{itemize}

\begin{figure}
    \centering
    \includegraphics[scale=0.5]{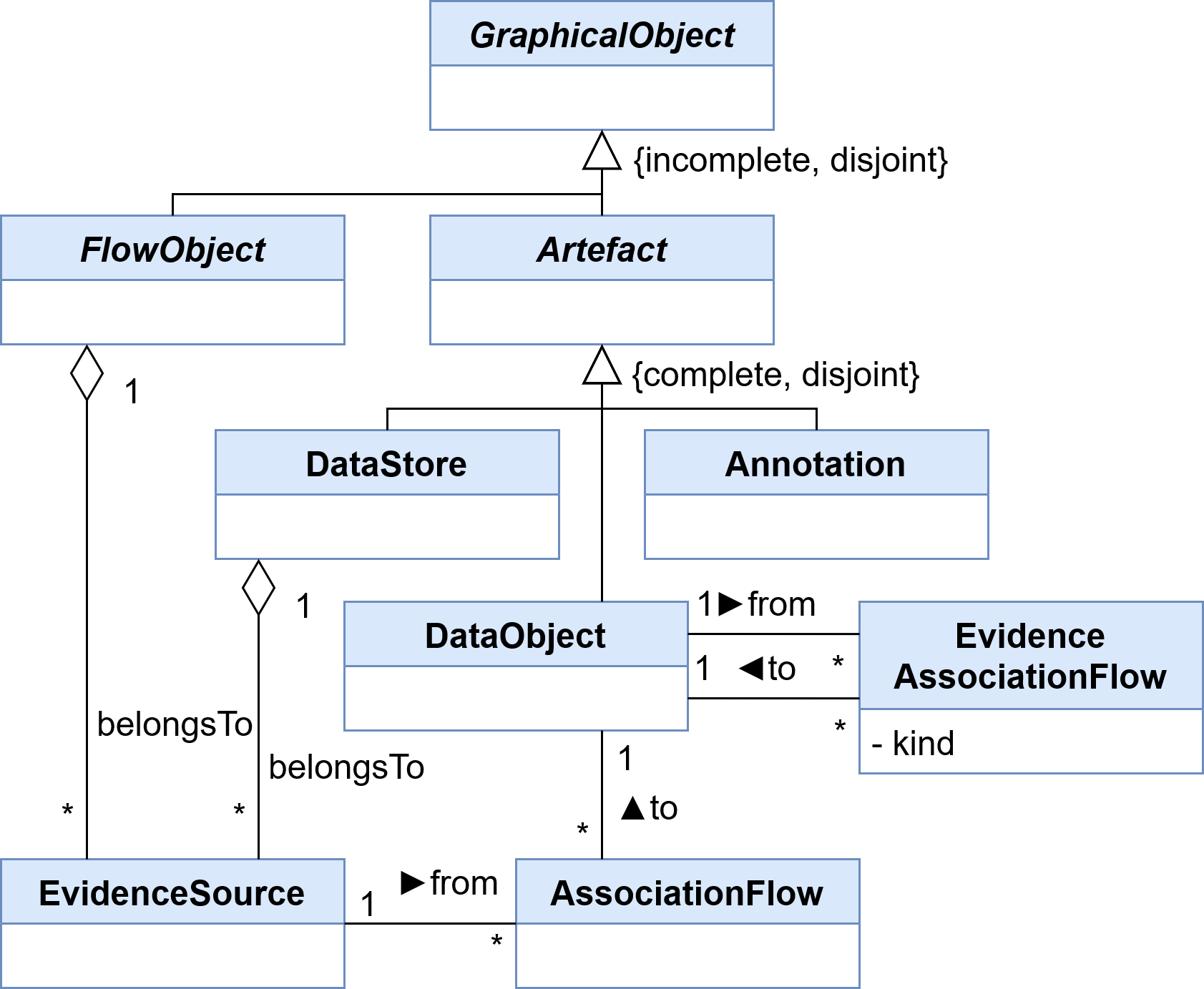}
    \caption{BPMN Syntax Excerpt with Forensic-Ready Modelling Notation}
    \label{fig:syntax}
\end{figure}

To satisfy the presented points, we introduce a new visual element, the \textit{Evidence Source}, visualized as a magnifying glass, used to represent the potential evidence source and bundle all its components. Together with the Evidence Source, we also introduce the \textit{Evidence Association}, representing different relationships between the potential evidence source data. Figure~\ref{fig:syntax} depicts the syntactical relation of the elements, and Table~\ref{tab:constructs} list their basic constructs together with their semantics.

\begin{table}
\caption{Forensic-Ready BPMN Constructs}
\label{tab:constructs}
\centering
\scriptsize
\begin{tabular}{| >{\centering\arraybackslash} m{0.25\linewidth} | >{\raggedright\arraybackslash} m{0.3\linewidth} | >{\raggedright\arraybackslash} m{0.3\linewidth} |}
\hline
Construct & \centering\arraybackslash Syntax & \centering\arraybackslash Semantics \\
\hline
\includegraphics[scale=0.44]{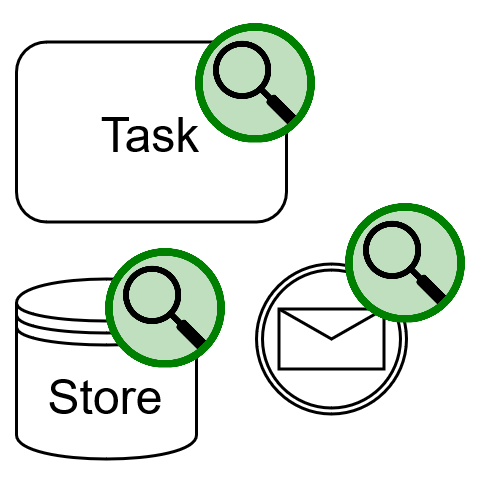} & EvidenceSource belonging to FlowObject or DataStore & IS asset marked as potential evidence source part, with one or more type of data missing \\
\hline
\includegraphics[scale=0.44]{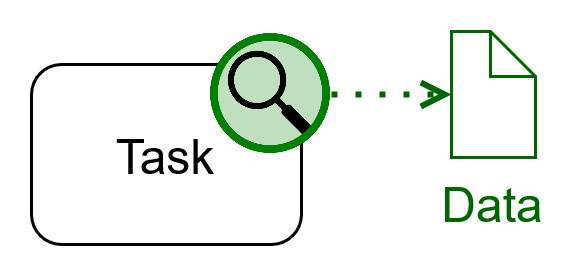} & AccociationFlow form EvidenceSource to DataObject & Potential evidence source with one type of data \\
\hline
\includegraphics[scale=0.44]{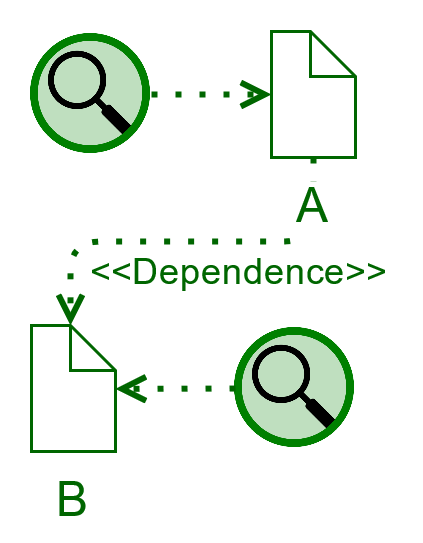} & EvidenceAssociationFlow with Dependence kind from DataObject A to DataObject B, both associated with EvidenceSource & If process execution is nominal, then before an occurrence of potential evidence B, there must be an occurrence of potential evidence A \\
\hline
\includegraphics[scale=0.44]{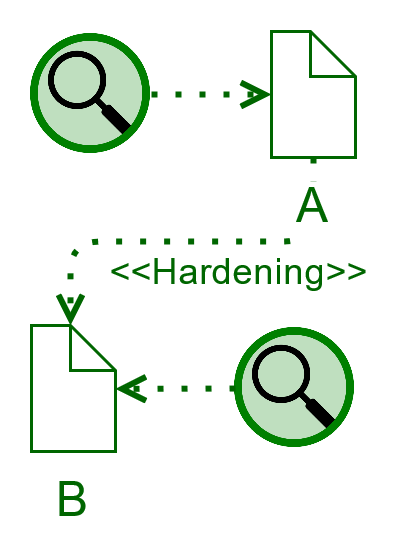} & EvidenceAssociationFlow with Hardening kind from DataObject A to DataObject B, both associated with EvidenceSource & Potential evidence A is strengthening potential evidence B by providing additional proof (e.g., integrity, authenticity) \\
\hline
\end{tabular}
\end{table}

Fundamentally, the notation enables the modelling of the forensic-ready requirements formulated by the risk analysis process. The Evidence Source – Data Object tuple represents the refined forensic-ready requirement within the system. In other terms, the Evidence Source represents a point of origin of the potential evidence. In contrast the (possibly multiple) associated Data Objects represents the types of data consisting of the potential evidence in question. Meaning, the concrete pieces of potential evidence are the instances of Data Objects produced by the process executions. Any relations between such Data Objects denotes a further semantic relationship, beneficial for further evaluation and documentation of forensic readiness.

\section{Approach Illustration}
\label{sec:demo}

\begin{figure*}
    \centering
    \begin{minipage}{\linewidth}
        \begin{center}
        \includegraphics[scale=.33]{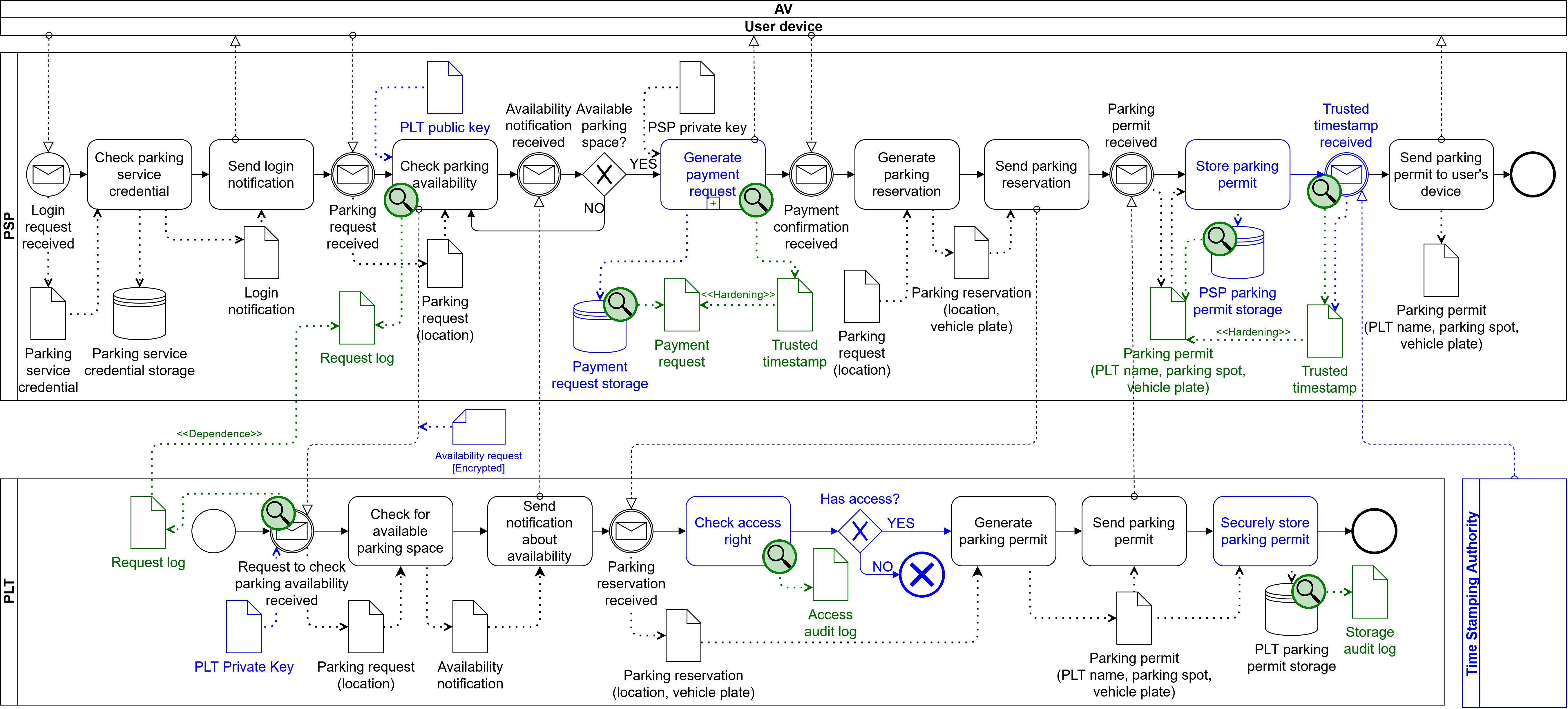}
        \end{center}
    \end{minipage}
    \caption{Automated Valet Parking Scenario --- Forensic-Ready Design}
    \label{fig:scenarioFinal}
\end{figure*}

In this section, we illustrate how to apply the extended notations to design a forensic-ready system on the AVP scenario. The forensic-ready risk analysis process starts after the security risk management process is finished. Therefore, the security risks in the running scenario are elicited, treated, and requirements formulated concerning the attack scenarios. The risks are listed in Tables~\ref{tab:case1}-\ref{tab:case6}, and the security risk treatment is modelled in Figure~\ref{fig:scenarioFinal} in blue colour, as mandated by the Security Risk-Aware BPMN. Furthermore, security patterns~\cite{Ahmed:2014, Matulevivcius:2017} have been applied to mitigate the risks in Cases \#1-3 further to demonstrate the complementary nature of the forensic-ready design.

\begin{table}
\caption{Case \#1: Covert Denial of Service}
\label{tab:case1}
\centering
\scriptsize
\begin{minipage}{\linewidth}
\begin{center}
\begin{tabular}{| >{\raggedright\arraybackslash} m{0.2\linewidth} | m{0.7\linewidth} |}
\hline
Risk & An attacker eavesdrops on the transmission medium, intercepts parking availability requests to PLT, and replying negatively on parking space availability at random times due to insecure protocol usage, leading to covert disruption of PLT availability, aborting the reservation process. \\
\hline
Security risk treatment & Risk Mitigated \\
\hline
Security requirement & Make the availability requests and responses unreadable to an attacker using public-key cryptography.$^*$ \\
\hline
Residual risk & An attacker reverse-engineers the PLT private key, decrypt the communication and resume the attack. \\
\hline
Forensic-ready requirement & Introduce request and response logging on both communication sides. A comparison of the logs shows modified requests. \\
\hline
\end{tabular}
\end{center}
\footnotesize
\quad\quad $^*$Corresponds to a security pattern SRP2~\cite{Matulevivcius:2017}.
\end{minipage}
\end{table}
\begin{table}
\caption{Case \#2: Parking Permit Fabrication (Store Injection)}
\label{tab:case2}
\centering
\scriptsize
\begin{minipage}{\linewidth}
\begin{center}
\begin{tabular}{| >{\raggedright\arraybackslash} m{0.2\linewidth} | m{0.7\linewidth} |}
\hline
Risk & A malicious insider injects a parking permit into the PLT parking permit storage out-of-the-process due to their ability to access it. Leading to a loss of parking permit integrity. \\
\hline
Security risk treatment & Risk Mitigated \\
\hline
Security requirement & Make the parking permit, inserted into PLT parking permit storage, invisible for other subjects.$^*$ \\
\hline
Residual risk & A malicious insider with elevated privileges may still access the data and resume the attack. \\
\hline
Forensic-ready requirement & Include data store auditing. Use the logs in monitoring to uncover suspicious behaviour. \\
\hline
\end{tabular}
\end{center}
\footnotesize
\quad\quad $^*$Corresponds to a security pattern SRP5~\cite{Matulevivcius:2017}.
\end{minipage}
\end{table}
\begin{table}
\caption{Case \#3: Parking Permit Fabrication (Access Control)}
\label{tab:case3}
\centering
\scriptsize
\begin{minipage}{\linewidth}
\begin{center}
\begin{tabular}{| >{\raggedright\arraybackslash} m{0.2\linewidth} | m{0.7\linewidth} |}
\hline
Risk & An attacker fabricates a fake Parking reservation and sends it to PLT due to missing access control. Leading to the loss of parking permit integrity. \\
\hline
Security risk treatment & Risk Mitigated \\
\hline
Security requirement & Check the access rights before processing the parking reservation request. Allow only the authorized requests.$^*$ \\
\hline
Residual risk & An attacker tampers the access control to authorize attacker requests and resume the attack. \\
\hline
Forensic-ready requirement & Include the access control auditing. Use the logs in monitoring to uncover suspicious behaviour. \\
\hline
\end{tabular}
\end{center}
\footnotesize
\quad\quad $^*$Corresponds to a security pattern SRP1~\cite{Matulevivcius:2017}.
\end{minipage}
\end{table}
\begin{table}
\caption{Case \#4: Issuing a Fake Payment Request}
\label{tab:case4}
\centering
\scriptsize
\begin{tabular}{| >{\raggedright\arraybackslash} m{0.2\linewidth} | m{0.7\linewidth} |}
\hline
Risk & An attacker uses a leaked PSP private key to impersonate PSP and creates a fake payment request, leading to loss of payment request integrity. \\
\hline
Security risk treatment & Risk Retained, leakage of PSP private key has a low probability. \\
\hline
Forensic-ready requirement & Retain the payment request history with proof of integrity and trusted timestamp. \\
\hline
\end{tabular}
\end{table}
\begin{table}
\caption{Case \#5: Dispute - Parking Permit Repudiation}
\label{tab:case5}
\centering
\scriptsize
\begin{tabular}{| >{\raggedright\arraybackslash} m{0.2\linewidth} | m{0.7\linewidth} |}
\hline
Risk & A dishonest customer repudiates a Parking permit that they received from PSP and stored on a User device, therefore out of the organization control, demanding a reimbursement, leading to a financial loss. \\
\hline
Forensic-ready requirement & Retain the issued parking ticket in PSP with proof of integrity and trusted timestamp. \\
\hline
\end{tabular}
\end{table}
\begin{table}
\caption{Case \#6: Unknown Risk}
\label{tab:case6}
\centering
\scriptsize
\begin{tabular}{| >{\raggedright\arraybackslash} m{0.2\linewidth} | m{0.7\linewidth} |}
\hline
Risk & An attacker uses zero-day vulnerability to send a fake Parking reservation to PLT.  Leading to the loss of parking permit integrity. \\
\hline
Forensic-ready requirement & Introduce a logging of communication on the interface and auditing on data stores. \\
\hline
\end{tabular}
\end{table}

The initial step of the process is to identify the initial sources of potential evidence. It was already performed partially and informally when discussing the innate evidence sources presented in Figure~\ref{fig:scenarioBasic}, but all found insufficient. However, considering the refined security requirements, new sources of evidence might appear. Concretely, the security requirement of Case \#3 (see Table~\ref{tab:case3}) corresponding to a security pattern SRP5 (Securing data stored in/retrieved from the data store)~\cite{Matulevivcius:2017}, which includes monitoring of an audit log, essentially a (foreground) source of potential evidence.

In re-evaluating the results form after the security risk management, risk mitigation and retention decisions are considered. The goal is to find whether the taken risk treatment could be overridden or enhanced by considering the forensic-ready means. Firstly, the already mitigated risks, concretely Cases \#1-3 (see Tables~\ref{tab:case1}-\ref{tab:case3}), are examined for residual risks described in the corresponding tables. Secondly, the decision to retain the risk, like in Case \#4 (see Table~\ref{tab:case4}) for the low probability of occurrence, is reviewed. Even though the mitigating security requirement is deemed unnecessary, the forensic-ready requirement offers the means to register its event. Additionally, the currently unknown risk, demonstrated by Case \#6 (see Table~\ref{tab:case6}), is considered, although with inherently vague description, as a basis for further refinement of forensic-ready requirements.

The discussed attack scenarios feature two cases of dispute concerning business asset and potential evidence source. The former is Case \#5 (see Table~\ref{tab:case5}), describing the risk of a business dispute with a dishonest customer. While the latter, Case \#4 (see Table~\ref{tab:case4}), brings up another requirement on hardening potential evidence source. Both cases tackle the issue by providing a trusted timestamp~\cite{Cosic:2010}, simultaneously protection and precision measure. Formally, the disputes would consider an iteration after the introduction of the corresponding sources.

Finally, based on the formulated risks from the previous two steps, new sources of potential evidence are created or modified accordingly. It corresponds with the formulation of forensic-ready requirements and is supported by the notation. All the forensic-ready requirements for the discussed risks are listed in the corresponding Tables~\ref{tab:case1}-\ref{tab:case6}. Notation-wise, Figure~\ref{fig:scenarioFinal} contains the modelled sources of potential evidence in a graphical way, highlighted in green colour, except for Case \#6 omitted for clarity.

Furthermore, the dependence and hardening relationships between the potential evidence data are highlighted by the notation. An example of the former is on Case \#1, where a nominal execution must produce two entwined records, a Request log on the PSP side before a Request log on the PLT side. An abnormal performance (attack, malfunction) should be reflected by the existence, content, or ordering of records. The latter is demonstrated by Case \#4 and \#5, where a trusted timestamp serves as proof of integrity and timeliness.

\section{Discussion}
\label{sec:discussion}

The application of both process and notation within Risk-Oriented Forensic-Ready Design demonstrated a high-level method for formulating and modelling the forensic-ready properties of software systems. One of the important features is understandability for a security engineer with basic knowledge of digital forensics as the process builds on the well-known risk-oriented security approaches. The notation itself serves as an understandable view: (1) what to implement in the system and (2) where to find the potential evidence.

Our approach focuses on both the security risks and business disputes, as evident from the demonstration. The security aspect is the most obvious one, important in detecting and tracing both attacks from outside and inside. Therefore, we motivate the consideration of the inner behaviour of the systems concerning those attacks as an addition to the classical frontier defence. Furthermore, we include the support for disputable assets both in terms of security and business. However, the approach does not explore the question of dependability, which the presented constructs could arguably support. The reasoning behind this is that one can utilize the potential evidence sources in discovering the root cause of failure~\cite{ErolKantarci:2013}.

Based on the list of requirements on forensic-ready software systems~\cite{Pasquale:2018}, the evidence availability (Evidence Source – Data Object tuple) and non-repudiation (Hardening Evidence Association) are explicitly considered in this paper. Arguably, the notation could be further used for data provenance requirements and linkability requirements. The former can be seen as a particular case of Evidence Association similar to Hardening. However, to fully include such support, its specifics need to be explored further. The latter is modelled by the Dependency Evidence Association, as both pieces must have a common identifier.

Additionally, it is implied by the process execution itself (at least partially following the Sequence Flow). Although, the details are left on a free-text description. Further exploration of the likability of potential evidence within process modelling could be utilized for analysis using process mining~\cite{vanderAalst:2016}.

\textbf{Model analysis} and possible automation were considered while developing the presented BPMN notation, although it is not explored in this paper. The idea is fourfold: (1) verify the correctness of the design according to the rules, (2) give hints during the design process, (3) provide insights into the potential evidence originating within the system, and (4) verify the implementation based on the model. The following paragraphs provide examples within the demonstrated model.

For design-time, rule checking, Case \#1 contains a good example of such a rule, where a Request log needs to be on both communicating parties. The reason is visible from the corresponding attack model in Figure~\ref{fig:attack1}. If either of those Request logs would be missing, then it would be impossible to detect the attack. Generally, the presence, ordering, and contents (i.e., alignment) of potential evidence within a process execution should serve as an indicator of compromise. The rules and constructs like the Dependency Evidence Association should help to ensure such properties.

The rules can also be applied in a more relaxed manner to give hints during the modelling. These can be understood as guides or warnings. For example, when using Hardening Evidence Association, it might be advisable to have the two Data Objects in separate Pools to decrease tampering risk. Although it ultimately depends on the data and its context.

Beyond the design, there is a possibility to use the models for documentation. Related potential evidence can be highlighted, showing the vital identifiers and how they fit into the context, giving valuable insights. We propose this view for the investigators to serve as documentation attached to the potential evidence data.

Another type of verification is focused on run-time, where the process mining~\cite{vanderAalst:2016} could be utilized. In this case, the potential evidence produced by the process execution would serve as an event log. Then the conformance checking technique could potentially display deviations from the model. Alternatively, one could obtain a model from the event log with process discovery. Furthermore, the utilization of process mining would imply linkability of potential evidence.

\section{Conclusion}
\label{sec:conclusion}

We presented a risk-oriented design approach for forensic-ready software systems, which was also practically demonstrated in a running scenario. The proposal consists of two components including the process for eliciting the specific forensic-ready requirements and the BPMN extension to present them. Notably, the process builds on security risk management to assess risks and, based on them, formulate the requirements. The notation supports this process and, most importantly, allows high-level modelling of potential evidence sources for their implementation within the system. Finally, we discussed the contributions of our approach and outlined the possibilities of further evolution. Our proposal could support design of the forensic readiness into the software systems to improve the overall security of IT infrastructure.
\begin{acks}
 This research was supported by ERDF "CyberSecurity, CyberCrime and Critical Information Infrastructures Center of Excellence" (No. CZ.02.1.01/0.0/0.0/\allowbreak16\_019/0000822) and EU Horizon 2020 research and innovation programme under grant agreement No 830892, project SPARTA.
\end{acks}

\bibliographystyle{ACM-Reference-Format}
\bibliography{bibliography}


\end{document}